# Long Tail Behavior of Queue Lengths in Broadband Networks: Tsallis Entropy Framework

Karmeshu[1] and Shachi Sharma[1,2]

[1] #School of Computer & Systems Sciences,

Jawaharlal Nehru University, New Delhi-110067, India

[2] Centre for Development of Telematics, New Delhi- 110030, India

\# Corresponding Author: Prof. Karmeshu

Email: *karmeshu@mail.jnu.ac.in*, *karmeshu_jnu@yahoo.com*

Phone: +91-11-26704767, Fax: +91-11-26717526

# Long Tail Behavior of Queue Lengths in Broadband Networks: Tsallis Entropy Framework

**Abstract:** A maximum entropy framework based on Tsallis entropy is proposed to depict long tail behavior of queue lengths in broadband networks. Queue length expression as measured in terms of number of packets involves Hurwitz-zeta function. When the entropy parameter $q$ in Tsallis entropy is less than unity, the distribution of packets yields power law behavior. In the limit $q$ tending to unity, Tsallis entropy expression reduces to one due to Shannon and well-known results of M/M/1 queuing system are recovered. Relationship between Tsallis entropy parameter $q$ and Hurst parameter $H$ (measure of self-similarity) is postulated. A numerical procedure based on Newton-Raphson method is outlined to compute Lagrange's parameter $\beta$. Various relationships between traffic intensity $\rho$ and Lagrange's parameter $\beta$ are examined using data generated from mean number of packets from storage model due to Norros. It is found that best fit corresponds to $\rho$ being a linear combination of decaying exponential and power exponent in $\beta$ for different values of entropy parameter $q$. Explicit expression for the probability that queue size exceeds a certain value is derived and it is established that it asymptotically follows power law for $q$ less than one. The system utilization shows an interesting behavior when the parameter $\rho$ is varied. It attains lower values than that of M/M/1 system for smaller values of $\rho$ whereas situation reverses for higher values of $\rho$.

**Keywords:** Long range dependence, Tsallis entropy, Maximum entropy Principle, Overflow probability, Utilization

## 1. Introduction

Extensive research on nature of traffic in local area network (LAN), wide area network (WAN), broadband networks (B-ISDN) and wireless packet networks has confirmed the presence of self-similarity and long range dependence[1-5] characterized by long tail behavior. Self-similar feature of traffic has been modelled in terms of Fractional Brownian Motion (FBM) process with Hurst parameter H lying in the range $0.5 < H < 1$. Employing FBM as an input process in queuing system, Norros [6] has proposed a simple storage model for ATM networks that gives an expression for calculating buffer capacity,

$$\overline{N} = \frac{\rho^{1/[2(1-H)]}}{(1-\rho)^{H/(1-H)}}, \qquad 0.5 < H < 1 \qquad (1)$$

A maximum entropy framework based on Tsallis entropy is proposed to capture long tail behavior. Tsallis rediscovered Havrda-Charvat measure of entropy and demonstrated the capability of this entropy measure to capture power like distributions in statistical mechanics with non-extensive properties. This entropy has been successfully employed to a variety of systems having multifractal structures [7]. Recently, Suyari [8] has employed maximum entropy principle to Tsallis entropy to study $q$-Gaussian distribution. The maximum entropy formalism due to Jaynes [9] has earlier been employed to analyse queueing problems by maximizing Shannon entropy when moment constraints are given [10-13].

It is proposed to employ maximum entropy principle based on Tsallis entropy to queueing system for gaining more insight into various quality of service (QoS) parameter. The paper contains five sections. Section 2 discusses the Tsallis entropy and maximum entropy framework to yield probability density of queue sizes measured in terms of number of packets. Section 3 provides a numerical procedure to estimate the Lagrange's parameter and also analyses the relatiopnship between $\rho$ and $\beta$. Section 4 deals with overflow probability and other performance measures. The last section 5 contains concluding remarks.

## 2. Tsallis Entropy and Queue Length Distribution

Tsallis [14, 15] proposed a parametric entropy measure which can be viewed as a generalization of Shannon entropy to deal with non-extensive systems. In terms of probability $p_i$ of the system being in state $i$, Tsallis defined the entropy function,

$$S_q = K \left(1 - \sum_{i=0}^{\infty} p_i^q \right) \bigg/ (q-1) \qquad (2)$$

where the parameter $q$ measures the degree of nonextensivity of the system. It is easy to see that Tsallis entropy reduces to Shannon entropy as parameter $q$ approches unity.

We employ the MEP framework to study queue length distribution of number of packets in the network when prior information in the form of mean number of packets is available. The queuing problem can be formulated as

$$\text{Max } S_q = \text{Max } K \left(1 - \sum_{i=0}^{\infty} p_i^q \right) \bigg/ (q-1) \qquad (3a)$$

subject to

$$\sum_{i=0}^{\infty} i \, p_i = A \qquad (3b)$$

$$\sum_{i=0}^{\infty} p_i = 1 \qquad (3c)$$

where *A* signifies mean queue size. Using Lagrange's method of undetermined multipliers, the Lagrangian function is given by,

$$\phi_q = \frac{S_q}{K} - \alpha(1 - \sum_{i=0}^{\infty} p_i) + \alpha\beta(q-1)(1 - \sum_{i=0}^{\infty} i\, p_i) \qquad (4)$$

where $\alpha$ and $\beta$ are Lagrange's multipliers. On imposing $\partial \phi_q / \partial p_i = 0$, we get

$$p_i = \left( [1 + \beta(1-q)i]^{1/(q-1)} \right) \Big/ \left( \sum_{i=0}^{\infty} [1 + \beta(1-q)i]^{1/(q-1)} \right), \quad i = 0, 1, 2, \ldots \qquad (5)$$

Noting that normalizaion constant in (5) remains finite, we require Re[1/(1-*q*)] > 1 i.e. *q* > 0. The probablity distribution for the queue length distribution as given by (5) can be rewritten as

$$p_i = \left( \left[ \frac{1}{\beta(1-q)} + i \right]^{1/(q-1)} \right) \Big/ \varsigma\left[ \frac{1}{1-q}, \frac{1}{\beta(1-q)} \right], \quad q > 0, \ i = 0, 1, 2, \ldots$$

(6)

where $\varsigma\left[ \frac{1}{1-q}, \frac{1}{\beta(1-q)} \right]$ denotes the Hurwitz-Zeta function [16] defined by

$$\varsigma\left[ \frac{1}{1-q}, \frac{1}{\beta(1-q)} \right] = \sum_{i=0}^{\infty} \left[ i + \frac{1}{\beta(1-q)} \right]^{-1/(1-q)} \qquad (7)$$

The probability distribution of $p_i$ is also known as the Zipf-Mandelbrot distribution [17, 18]. The parameter $\beta$ in (6) can be estimated from the constraints given in (3b),

$$A = \left( \varsigma\left[ \frac{q}{1-q}, \frac{1}{\beta(1-q)} \right] \Big/ \varsigma\left[ \frac{1}{1-q}, \frac{1}{\beta(1-q)} \right] \right) - \frac{1}{\beta(1-q)}, \ \text{Re}[q/(1-q)] > 1 \ \text{ or } \ q > 1/2 \qquad (8)$$

It is easy to see that in the limit $q \to 1$, (5) reduces to

$$p_i = e^{-\beta i} \bigg/ \left( \sum_{i=0}^{\infty} e^{-\beta i} \right), \qquad i = 0, 1, 2, \ldots \qquad (9)$$

which is well-known expression of probability distribution in steady state for M/M/1 queue system with traffic intensity $\rho = e^{-\beta}$. However, for $q < 1$, we find from (5) that for large number of packets $i$, $p_i$ follows power law,

$$p_i \sim i^{1/q-1}, \quad q < 1 \qquad (10)$$

From (8) and (10), it can be infered that $\frac{1}{2} < q < 1$.

Note that for $H = 1/2$, (1) yields result for mean queue size of M/M/1 system which corresponds to the case $q = 1$ in the proposed framework. We postulate a relationship between Tsallis entropy parameter $q$ and Hurst parameter $H$,

$$q = 1.5 - H$$

(11)

For evaluation of QoS parameters, we require to explore relationship between traffic intensity $\rho$ and Largarnge's parameter $\beta$. The next section deals with this aspect.

## 3. Exploring Relationships between $\rho$ and $\beta$

In order to explore relationship between $\rho$ and $\beta$, we need to estimate first the Largarnge's parameter $\beta$ from the constraint (3b). We extend the method proposed by Johnson [19] in the context of Shannon entropy. The constraints (3b) can be expressed as

$$\sum_i (i-A) p_i = 0 \tag{12}$$

Using (5), (12) can be rewritten as

$$\sum_i (i-A)[1+\beta(1-q)i]^{1/(q-1)} = 0 \tag{13}$$

Equation (13) is of form $f(\beta)=0$ and can be solved through Newton-Raphson method from an initial approximation of $\beta$ and replacing it at each iteration by $\beta+\Delta\beta$ where

$$\Delta\beta = \frac{-f(\beta)}{f'(\beta)} \tag{14}$$

This can be expressed as,

$$\Delta\beta = \frac{\beta(1-q)\left\{\varsigma\left[\frac{q}{1-q},\frac{1}{\beta(1-q)}\right] - \left(\frac{1}{\beta(1-q)} + A\right)\varsigma\left[\frac{1}{1-q},\frac{1}{\beta(1-q)}\right]\right\}}{\varsigma\left[\frac{q}{1-q},\frac{1}{\beta(1-q)}\right] - \left(\frac{2}{\beta(1-q)} + A\right)\varsigma\left[\frac{1}{1-q},\frac{1}{\beta(1-q)}\right] + \frac{1}{\beta(1-q)}\left(A + \frac{1}{\beta(1-q)}\right)\varsigma\left[\frac{2-q}{1-q},\frac{1}{\beta(1-q)}\right]} \tag{15}$$

till the sequence of iterates converges.

In order to derive the relationship between $\rho$ and $\beta$, we assume that $\rho$ is a function of $\beta$ and $q$ i.e.

$$\rho = f(\beta, q) \tag{16}$$

For estimating $f(.)$, we rewrite (1) in the form

$$\rho^{*2H} Y + \rho^* = 1 \tag{17}$$

where $\rho^* = 1-\rho$ and $Y = \overline{N}^{2(1-H)}$. The mean queue size as obtained from the maximization of Tsallis entropy (8) is compared with the one due to Norros (1) to enable calculations of $\rho$ and $\beta$ for a given value of mean queue size. We now outline the algorithm for estimating $\beta$ and $\rho$.

**Algorithm:**

**Input Data**

$q$, A and initial approximation of $\beta$

**Begin**

Step 1: Calculate $\Delta\beta$ as given in (15)

Step 2: $\beta \leftarrow \beta + \Delta\beta$

Step 3: Return to Step 1 until convergence of $\beta$

Step 4: $H = 1.5 - q$ and $\overline{N} \leftarrow A$

Step 5: Compute $\rho^*$ by applying (16)

Step 6: $\rho \leftarrow 1 - \rho^*$

**End**

**Output**

1. The Lagrange's parameter $\beta$

2. The parameter $\rho$

The plot between $\rho$ and $\beta$ for different values of $H$ or $q$ is shown in Figure 1. It is noted from the graph that $\rho$ and $\beta$ are inversely related. It is also clear when $q$ is in the neighbourhood of 1, $\rho$ is found to decay exponentially with respect to $\beta$ and when $q$ deviates from unity, a long tail behavior occurs. We now investigate different relationships between $\rho$ and $\beta$.

### 3.1 Modified Exponential Relation (Model I)

Motivated by the limiting case of $q$ tending 1, we attempt to see the validity of modified exponential function

$$\rho = a + b\,e^{-\beta} \tag{18}$$

where *a* and *b* are two parameters that are to be estimated. Using SPSS, the parameters *a* and *b* are estimated from the generated dataset by employing the algorithm outlined in previous subsection. The comparison of generated and original values of $\rho$ is depicted in Figure 2. It is observed that the fit gives good approximations only for lower values of $\beta$ when *q* is greater than 0.8. For lower values of *q*, there is significant deviation between original and generated values of $\rho$ resulting in poor fit.

**3.2 Combination of Exponential and Power Law (Model II)**

We now generalize above model as a linear combination of power exponent in $\beta$ and decaying exponential such that

$$\rho = c\,\beta^{-\eta} + d\,e^{-\mu\beta} \tag{19}$$

where *c*, $\eta$, *d* and $\mu$ are the parameters which needs to be determined utilizing the same datasets as used for (18). The graph between generated and original values of $\rho$ is shown also in Figure 2. It is found that the genrated values are very much closer to the original values in comparison to modified exponential relation. The non linear relationship as a combination of power exponent and exponential provides a better approximation than model I.

**4. Performance Measures**

Based on the foregoing analysis, we have been able to completely specify the probability distribution of queue size which can be used to provide quality of service (QoS) parameters.

**Mean and Variance of Number of Packets**

The mean number of packets is specified and can be expressed in terms of Hurwitz-zeta fucntion as,

$$\overline{N} = A = \left( \varsigma\left[ \frac{q}{1-q}, \frac{1}{\beta(1-q)} \right] \middle/ \varsigma\left[ \frac{1}{1-q}, \frac{1}{\beta(1-q)} \right] \right) - \frac{1}{\beta(1-q)} \quad (20)$$

The proposed framework has enabled us to derive the explicit expression of the varaince due to availiability of explicit form of probability distribution of the number of packets. The variance of the number of packets, defined by

$$\sigma_N^2 = \sum_{i=0}^{\infty} i^2 p_i - \overline{N}^2 \quad (21)$$

can be calculated using (6). This gives,

$$\sigma_N^2 = \left( \varsigma\left[ \frac{2q-1}{1-q}, \frac{1}{\beta(1-q)} \right] \middle/ \varsigma\left[ \frac{1}{1-q}, \frac{1}{\beta(1-q)} \right] \right) - \left( \varsigma\left[ \frac{q}{1-q}, \frac{1}{\beta(1-q)} \right] \middle/ \varsigma\left[ \frac{1}{1-q}, \frac{1}{\beta(1-q)} \right] \right)^2 \quad (22)$$

The numerator in (22) is finite only when $\text{Re}(2q-1)/(1-q) > 1$ or $(q > 2/3)$. The range of $q$ for variance to exist is $0.66 < q < 1$. This result can be generalized to $k^{th}$ order moment. The range of $q$ for the existence of $k^{th}$ order moment becomes $k/(1+k) < q < 1$ which shrinks as $k$ increases.

The plot of the variance with respect to parameter $\beta$ is depicted in Figure 3. We can observe that the variability of the number of packets increases very fast for much lower values of $\rho$ in contrast to M/M/1 queueing model where the sharp increase occurs in the vicinity of $\rho \approx 1$. Hence, as $q$ tends to ½, one encounters more bursty traffic leading to increased variance in the number of packets.

**Overflow probability**

A quantity of interest in networks is buffer overflow probability which is defined as,

$$P(i > x) = 1 - P(i \leq x) \quad (23)$$

From (6), we find,

$$P(i > x) = 1 - \sum_{i=0}^{x} \left[ \left[ \frac{1}{\beta(1-q)} + i \right]^{1/q-1} \right] \bigg/ \varsigma\left[ \frac{1}{1-q}, \frac{1}{\beta(1-q)} \right] \tag{24}$$

Using properties of Hurwitz-Zeta functions [16, pp-269],

$$P(i > x) = \left( 1 \bigg/ \varsigma\left[ \frac{1}{1-q}, \frac{1}{\beta(1-q)} \right] \right) \left( \frac{1-q}{q} \right) \left[ x + \frac{1}{\beta(1-q)} \right]^{-q/(1-q)}$$
$$- \left( 1 \bigg/ \varsigma\left[ \frac{1}{1-q}, \frac{1}{\beta(1-q)} \right] \right) \left( \frac{1}{1-q} \right) \sum_{n=x}^{\infty} \int_0^1 u \cdot \left( u + n + \frac{1}{\beta(1-q)} \right)^{-(2-q)/(1-q)} du \tag{25}$$

For asymptotically large $x$, we find the emergence of power law given as

$$P(i > x) \sim B \, x^{-q/(1-q)} \tag{26}$$

where

$$B = \left( 1 \bigg/ \varsigma\left[ \frac{1}{1-q}, \frac{1}{\beta(1-q)} \right] \right) \left( \frac{1-q}{q} \right)$$

(27)

It is clear from (26) that the distribution of the buffer content or queue length has long tail, thus implying that large buffers are required to cater to the bursty traffic. This is in conformity with the well known results obtained in context of broadband networks [20].

**Limiting Case as** $q \to 1$**:**

The overflow probability given by (24) can be rewritten as,

$$P(i > x) = \frac{1}{C}\left( \frac{1}{\beta q} \right) [1 + \beta(1-q)x]^{-q/(1-q)} - \frac{1}{C} \sum_{n=x}^{\infty} \int_0^1 u \, [1 + \beta(1-q)(u+n)]^{(2-q)/(1-q)} du \tag{28}$$

where

$$C = \sum_{i=0}^{\infty} [1 + \beta(1-q)i]^{1/(q-1)} \tag{29}$$

It would be interesting to see that one recovers the well-known results of exponentially decaying probability from (28),

$$P(i > x) \sim e^{-\beta x}$$

(30)

which corresponds to M/M/1 queue in the limiting case as $q \to 1$,

Figure 4 depicts behavior of buffer overflow probability for different values of Tsallis entropy parameter $q$. The buffer overflow probability is significantly high even for to low values of $\rho$ (corresponding large values of $\beta$) as $q$ tends to 0.5. This means long range dependence results in requirement of large buffer even at low traffic.

**System Utilization**

Utilization of the queueing system is defined as the probability that system is non empty i.e.

$$U = 1 - p_0 = 1 - \frac{[\beta(1-q)]^{1/(1-q)}}{\varsigma\left[\frac{1}{1-q}, \frac{1}{\beta(1-q)}\right]}$$

(31)

The utilization function for different values of $q$ and $\rho$ is plotted in Figure 5. An interesting finding is that the utilization is lower than that of M/M/1 queue for smaller values of $\rho$ and the situation undergoes a transition leading sharply to high utilization as one crosses some threshold value of $\rho$. For lower values of $q$, the transition from low to high utilization is fairly sharp as parameter $\rho$ varies – a characteristic feature of long-range dependence in network traffic.

**5. Conclusion**

A maximum entropy framework based on Tsallis entropy provides insight into infinite buffer queuing system when the first moment is known. An interesting aspect of the formulation is that one is able to express the probability distribution of number of packets in terms of Hurwitz-Zeta function. A power law behavior is observed when Tsallis parameter $q$ lies between 1/2 and 1. The results of M/M/1 queuing system are recovered as $q$ approaches 1. It is found that the range of $q$ for existence of $k^{th}$ order moment starts shrinking with increasing $q$. A numerical procedure based on Newton-Raphson method is also provided to compute Lagrange's parameter and hence obtain queue length distribution of number of packets. The traffic intensity $\rho$ is found to have exponential relation with Lagrange's parameter $\beta$ when $q$ is close to 1 and power law relation when $q$ tends to 0.5. It has been observed that traffic intensity $\rho$ can be expressed in terms of linear combination of decaying exponential and power exponent in $\beta$. The variance of number of packets in the system also exhibits power law behavior. The probability of exceeding certain threshold value of number of packets in the system is explicitly obtained and for large threshold values, a power law behavior is observed. A notable result arising from the study is that utilization is significantly different from that of M/M/1 queue, and it is found that system utilization marks a transition from low to high value as the parameter $\rho$ crosses some threshold value.

**Acknowledgement**

The authors would like to thank Prof. S.K. Rangarajan for extremely useful suggestions.

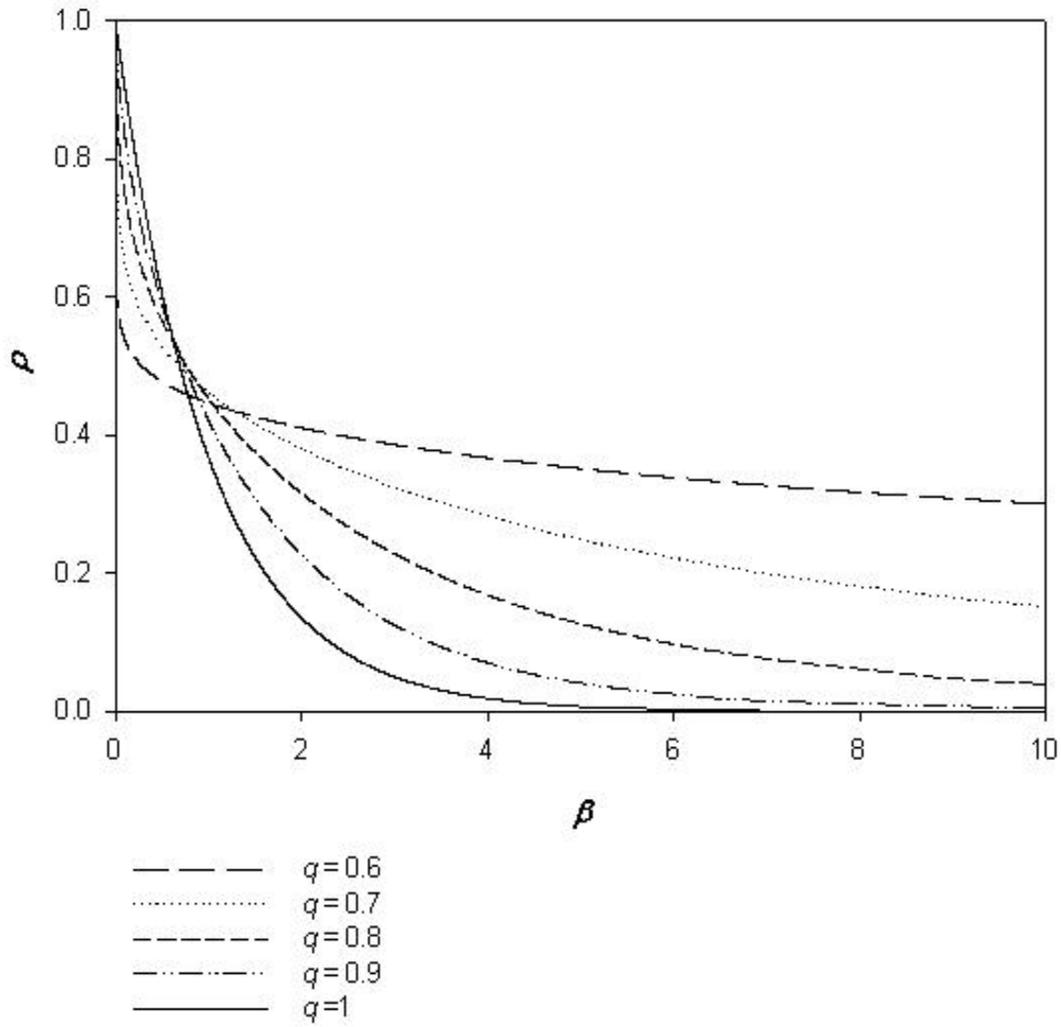

**Figure 1: Variation of $\rho$ with $\beta$ for different values of $q$**

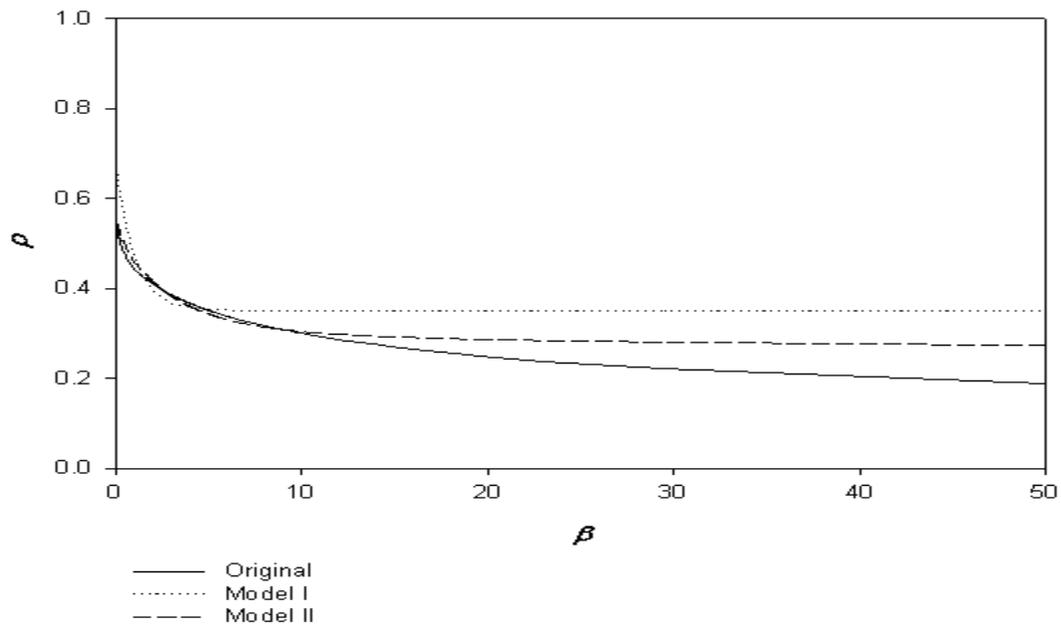

**(i)** *q*=0.6

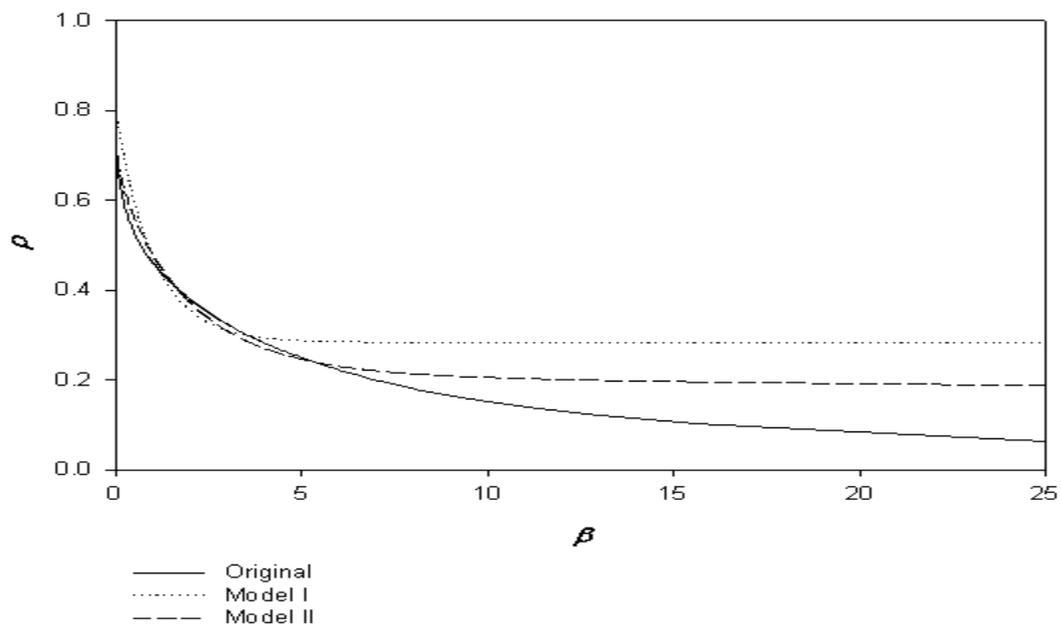

**(ii)** *q*=0.7

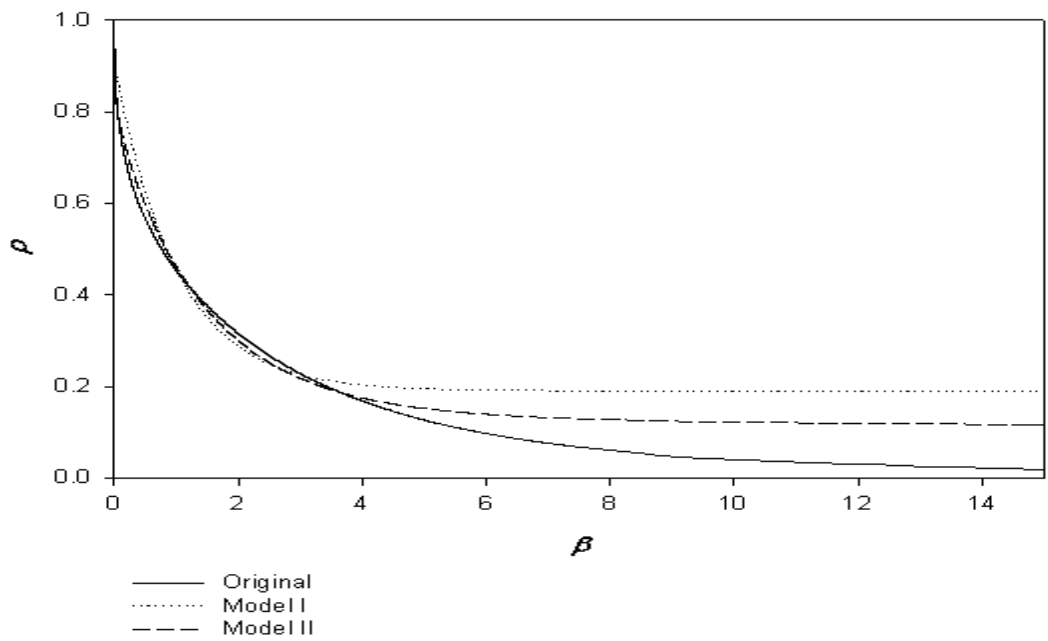

**(iii)** *q*=0.8

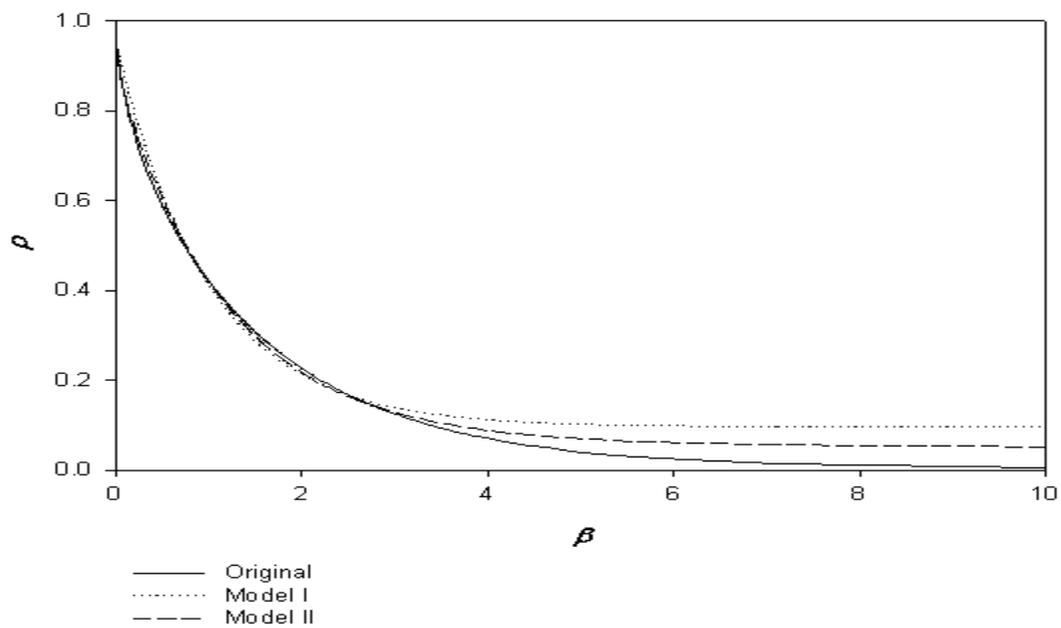

**(iv)** *q*=0.9

**Figure 2: Relationships between $\rho$ and $\beta$ for different values of *q***

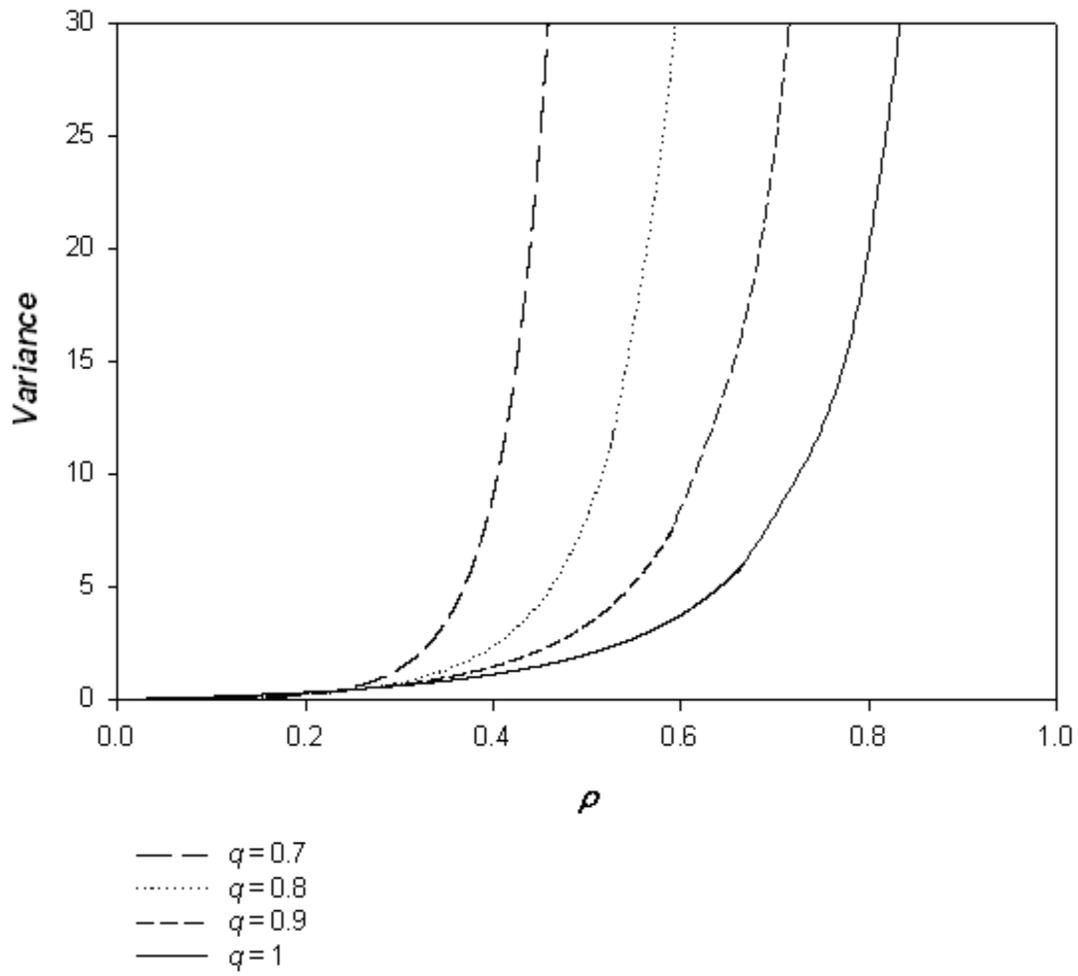

**Figure 3: Behavior of variance for number of packets with respect to $\rho$ for different parametric values of $q$**

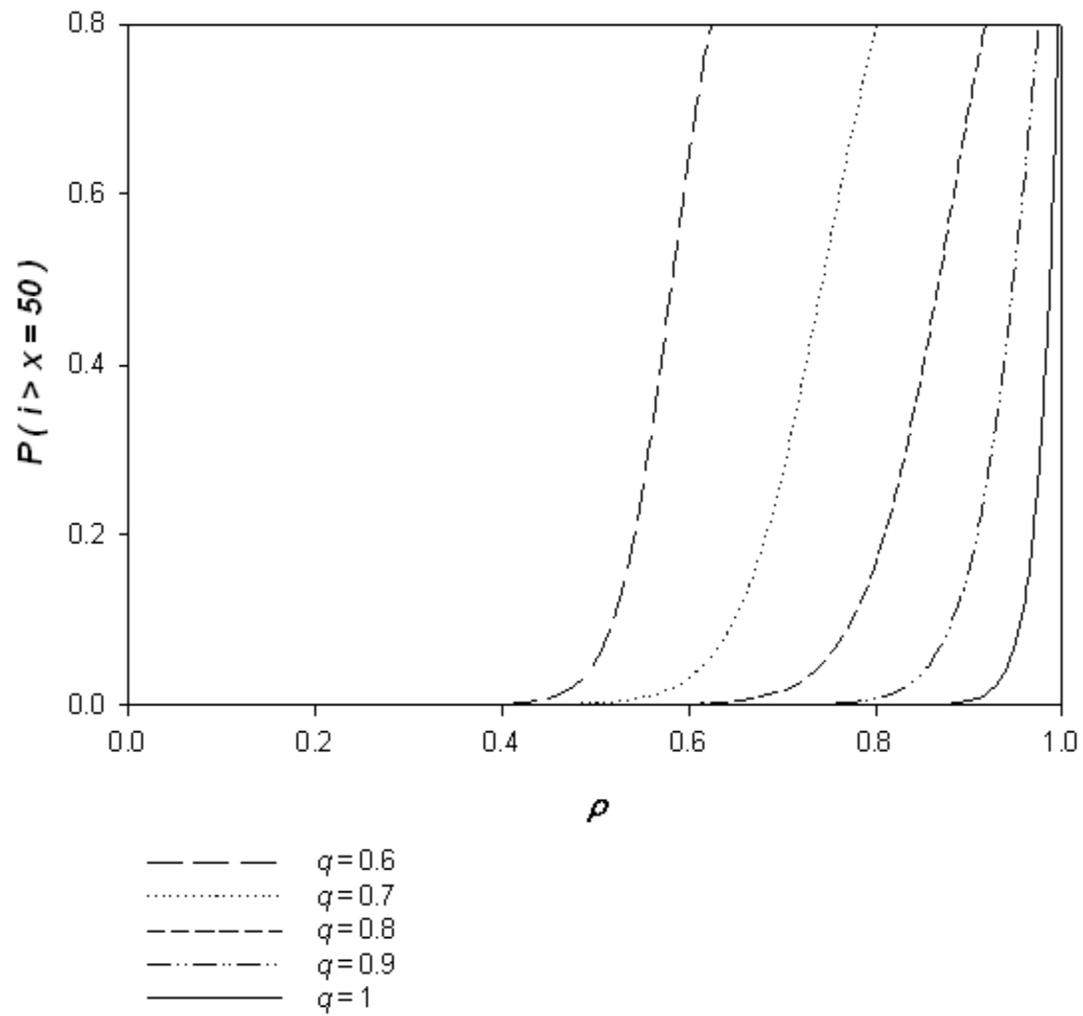

**Figure 4: Variation of Overflow Probability with respect to parameter $\rho$**

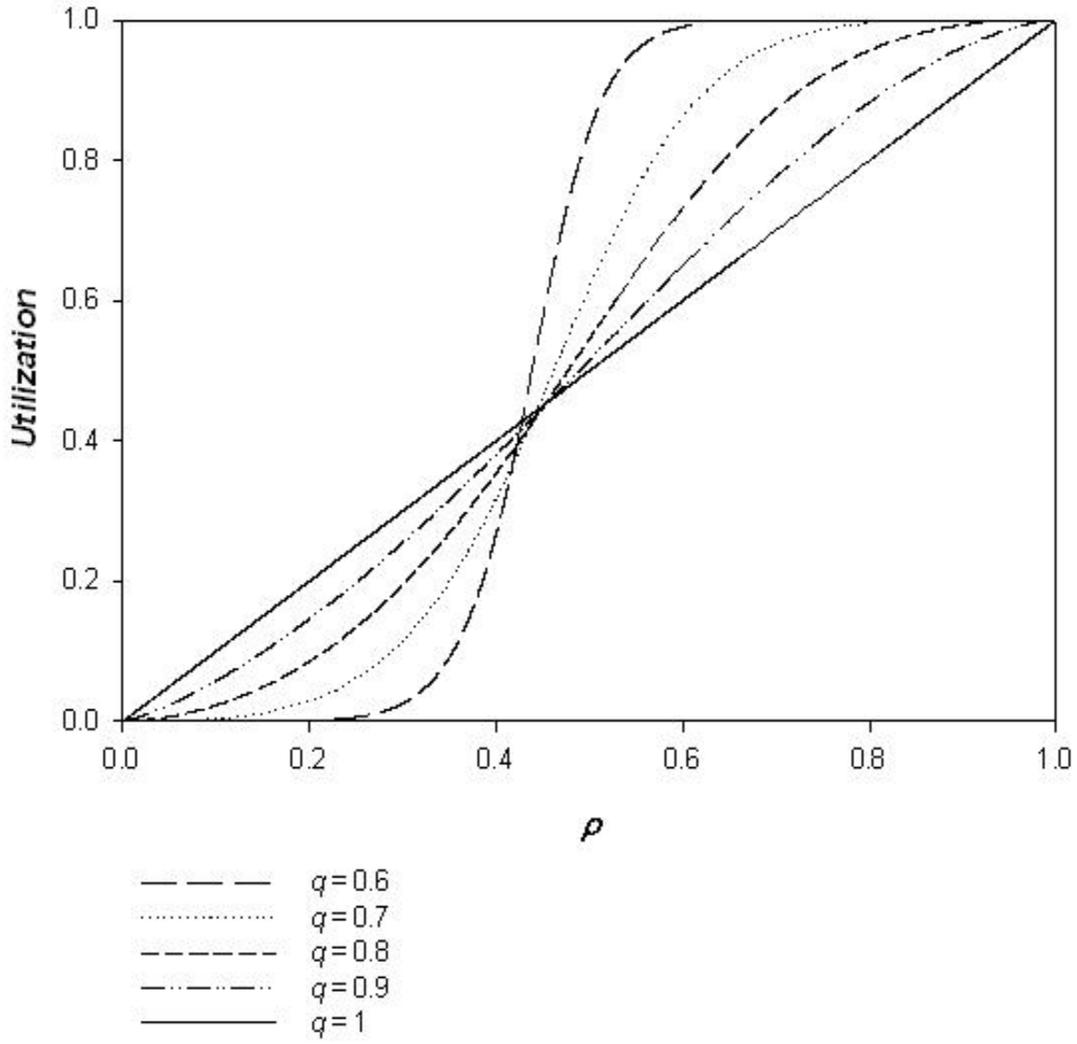

**Figure 5: Variation of queue Utilization with $\rho$**